\documentclass[12pt,onecolumn,aps,superscriptaddress,
               amsmath,amssymb,nofootinbib]{revtex4}      
\usepackage{graphicx}

\usepackage{graphicx,epsfig}

\usepackage{dcolumn}
\usepackage{bm}

\newcommand{\Z}{{\mathbb Z}}

\begin{document}

\begin{flushright}
\baselineskip=12pt \normalsize
{ACT-11-11},
{MIFPA-11-35}\\
\smallskip
\end{flushright}

\title{Proton Stability and Dark Matter in a Realistic String MSSM}

\author{James A. Maxin}
\affiliation{George P. and Cynthia W. Mitchell Institute for
Fundamental Physics, Texas A\&M
University,\\ College Station, TX 77843, USA}
\author{Van E. Mayes}
\affiliation{Physics Department, Arizona State University,\\ Tempe, AZ 85287-4111, USA}
\author{D.V. Nanopoulos}
\affiliation{George P. and Cynthia W. Mitchell Institute for
Fundamental Physics, Texas A\&M
University,\\ College Station, TX 77843, USA}
\affiliation{Astroparticle Physics Group, Houston
Advanced Research Center (HARC),
Mitchell Campus,
Woodlands, TX~77381, USA; \\
Academy of Athens,
Division of Natural Sciences, 28~Panepistimiou Avenue, Athens 10679,
Greece}

\begin{abstract}
\begin{center}
{\bf ABSTRACT}
\end{center}
We demonstrate the existence of an extra nonanomalous U(1) gauge symmetry in 
a three-generation Pati-Salam model constructed with intersecting D6-branes in Type IIA 
string theory on a $T^6/(\Z_2 \times \Z_2)$ orientifold.  This extra U(1) forbids
all dimension-4, 5, and 6 operators which mediate proton decay in the MSSM.  Moreover, this
results in the effective promotion of baryon and lepton number
to local gauge symmetries, which can potentially result in leptophobic and leptophilic 
$Z'$ bosons observable at the LHC.  
Furthermore, it is not necessary to invoke R-parity to forbid the dimension-4
operators which allow rapid proton decay.  However, R-parity may arise naturally from a spontaneously broken $U(1)_{B-L}$.  
Assuming the presence of R-parity, we then study the direct detection cross-sections for neutralino dark matter,
including the latest constraints from the XENON100 experiment. We find that these limits are now within required range necessary to begin testing the model.    
\end{abstract}

\maketitle

\newpage

\section{Introduction}

A main goal of string phenomenology is to discover the 
mechanisms by which the 
detailed properties of our universe may arise.  Among these
are the Standard Model (SM) gauge groups, the number of 
generations of chiral fermions, and the observed mass 
hierarchies and mixings of quarks and leptons.  
Of particular importance, the proton appears to have a very long lifetime.
Baryon ($B$) and lepton number ($L$) violating processes have to date never been observed, 
yet they are only conserved
as accidental global symmetries of the SM.  
However, such global symmetries
are generically broken by nonperturbative
effects, and thus baryon number is expected to be violated at some level
in the SM. 
In fact, dimension-4 operators appear in the minimal supersymmetric SM (MSSM)
which leads to proton decay at a disastrously high rate. Usually
these operators are eliminating by imposing a discrete symmetry on
the MSSM
such as R-parity~\cite{Fayet:1974pd,Fayet:1977yc, Nilles:1981py}.  
However, 
operators of dimension-5 also appear which are not eliminated by
R-parity, leading to a proton decay rate which is generically too large.  

Although imposition of R-parity may seem ad hoc, it provides a
simple explanation for another mystery.  Observations in cosmology
and astrophysics suggest the presence of a stable dark matter particle.
A natural candidate for WIMP-like dark matter is the lightest
supersymmetric partner (LSP)~\cite{Ellis:1983ew} in supersymmetric
models which include R-parity conservation, which is usually the
lightest neutralino $\widetilde{\chi}_{1}^{0}$~\cite{Ellis:1983ew, Goldberg:1983nd}.  
Limits on the dark matter relic abundance and direct and indirect detection 
cross-sections can be used to
constrain the possible superpartner and Higgs spectra which may
be observed at the Large Hadron Collider (LHC).  In short, only 
superpartner spectra which possess a stable LSP consistent with all other constraints 
on dark matter are viable.   

Of course, the MSSM is just an effective theory which should be 
replaced at high energies by something more fundamental, such as 
string theory.  In particular, Type IIA string compactifications 
involving $D6$-branes intersecting at angles (and their Type IIB 
duals including F-theory extensions) have provided a fruitful direction 
for studying this question.  Such models have been the subject 
of much study in recent years, and we refer the reader to
\cite{Blumenhagen:2005mu, Blumenhagen:2006ci} for recent reviews.   
A phenomenologically interesting model of this type 
was first constructed in~\cite{Cvetic:2004ui,Chen:2006gd} and studied 
in~\cite{Chen:2006gd, Chen:2007px, Chen:2007zu}.  In this 
three-generation Pati-Salam model, it is possible to obtain 
realistic Yukawa matrices for quarks and leptons, tree-level 
gauge unification at the string scale, and obtain realistic 
supersymmetry spectra satisfying all experimental constraints.  
The phenomenological consequences of this model at the LHC were 
considered in~\cite{Chen:2007zu} and~\cite{Maxin:2009ez}, and 
the implications for direct and indirect dark matter detection 
were initially studied in~\cite{Maxin:2009qq,Maxin:2009kp}.  
In the present 
work, we show that a variation of this model originally constructed
in~\cite{Chen:2007ms} possesses 
an extra nonanomalous U(1) gauge symmetry that forbids
all dimension-4, 5, and 6 operators found in the MSSM which 
allow proton decay (related  
four-generation models were considered in 
\cite{Belitsky:2010zr,Lebed:2011zg,Lebed:2011fw}).  
Thus, the proton is effectively stable in the model (see~\cite{Ellis:1997ec} for
a similar study in the context of free-fermionic 
heterotic string/M-theory compactifications).  
In particular,
it is not necessary to introduce R-parity in order to eliminate the 
dimension-4 operators which allow proton decay at a dangerously high rate.
Nevertheless, R-parity
may still naturally arise in the model via a
${\rm U}(1)_{B-L}$ gauge symmetry which is broken spontaneously
to its discrete $\Z_2$ subgroup, resulting in a stable LSP. 
Thus, we update the constraints on dark matter direct 
detection taking into account the recent limit on the dark matter 
direct detection cross-section from the CDMSII~\cite{Ahmed:2009zw}
and XENON100~\cite{Aprile:2011hx} collaborations.

\section{A Realistic MSSM with an Extra U(1)}

\begin{table}[t]
\caption{General spectrum for intersecting D6 branes at generic
angles, where $I_{aa'}=-2^{3-k}\prod_{i=1}^3(n_a^il_a^i)$ and
$I_{aO6}=2^{3-k}(-l_a^1l_a^2l_a^3
+l_a^1n_a^2n_a^3+n_a^1l_a^2n_a^3+n_a^1n_a^2l_a^3)$, where
$k = \beta_1+\beta_2+\beta_3$.  In addition,
${\cal M}$ is the multiplicity, and $a_S$ and $a_A$ denote
the symmetric and antisymmetric representations of
U($N_a/2$), respectively.}
\renewcommand{\arraystretch}{1.4}
\begin{center}
\begin{tabular}{|c|c|}
\hline {\bf Sector} & \phantom{more space inside this box}{\bf
Representation}
\phantom{more space inside this box} \\
\hline\hline
$aa$   & ${\rm U}(N_a/2)$ vector multiplet and 3 adjoint chiral
multiplets  \\
\hline $ab+ba$   & $ {\cal M}(\frac{N_a}{2},
\frac{\overline{N_b}}{2})=
I_{ab}=2^{-k}\prod_{i=1}^3(n_a^il_b^i-n_b^il_a^i)$ \\
\hline $ab'+b'a$ & $ {\cal M}(\frac{N_a}{2},
\frac{N_b}{2})=I_{ab'}=-2^{-k}\prod_{i=1}^3(n_{a}^il_b^i+n_b^il_a^i)$ \\
\hline $aa'+a'a$ &  ${\cal M} (a_S)= \frac 12 (I_{aa'} - \frac 12
I_{aO6})$~;~~ ${\cal M} (a_A)=
\frac 12 (I_{aa'} + \frac 12 I_{aO6}) $ \\
\hline
\end{tabular}
\end{center}
\label{spectrum}
\end{table}

Type IIA orientifold string compactifications with intersecting 
D-branes (and their Type IIB duals with magnetized D-branes) 
have provided exciting geometric tools with which the MSSM may
be engineered.  While this approach may not allow a
first-principles understanding of why the SM gauge groups and
associated matter content arises, it may allow a deeper insight into
how the finer phenomenological details of the SM may emerge.
In short, D6-branes in Type IIA fill
(3+1)-dimensional Minkowski spacetime and wrap 3-cycles in the
compactified manifold, such that a stack of $N$ branes generates a
gauge group U($N$) [or U($N/2$) in the case of $T^6/(\Z_2 \times
\Z_2)$] in its world volume.  On $T^6/(\Z_2 \times \Z_2)$, the
3-cycles are of the form~\cite{CSU}
\begin{equation}
\Pi_a = \prod_{i=1}^3(n_a^i[a_i] + 2^{-\beta_i} l_a^i[b_i]),
\end{equation}
where the integers $n_a^i$ and $l_a^i$ are the wrapping 
numbers around the basis cycles $[a_i]$ and $[b_i]$ of 
the $i$th two-torus, and $\beta_i=0$ for an untilted two-torus
while $\beta_i = 1$ for a tilted two-torus.    
In addition, we must introduce the orientifold images
of each D6-brane, which wraps a cycle given by
\begin{equation}
\Pi_a' = \prod_{i=1}^3(n_a^i[a_i] - 2^{-\beta_i} l_a^i[b_i]).
\end{equation}

In general, the 3-cycles wrapped by the stacks of D6-branes intersect
multiple times in the internal space, resulting
in a chiral fermion in the bifundamental representation localized at
the intersection between different stacks $a$ and $b$.  The multiplicity of such
fermions is then given by the number of times the 3-cycles intersect.
Each stack of D6-branes $a$ may 
intersect the orientifold images of other stacks $b'$, also resulting in fermions in
bifundamental representations.  Each stack may also intersect its own
image $a'$, resulting in chiral fermions in the symmetric and
antisymmetric representations.  The different types of representations
that may be obtained for each type of intersection and their
multiplicities are summarized in Table~\ref{spectrum}.  In addition, the
consistency of the model requires certain constraints to be satisfied,
namely, Ramond-Ramond (R-R) tadpole cancellation and the preservation 
of $\mathcal{N}=1$ supersymmetry.

\begin{table}[tf]
\footnotesize
\renewcommand{\arraystretch}{1.0}
\caption{D6-brane configurations and intersection numbers for
a three-family Pati-Salam model on a Type-IIA $T^6 / (\Z_2
\times \Z_2)$ orientifold, with a tilted third two-torus.  The complete gauge
symmetry is $[{\rm U}(4)_C \times {\rm U}(2)_L \times {\rm
U}(2)_R]_{\rm observable} \times [ {\rm U}(2) \times {\rm USp}(2)^2]_{\rm hidden}$ and
$\mathcal{N}=1$ supersymmetry is preserved for $\chi_1=3$, $\chi_2=1$, $\chi_3=2$.}
\label{MI-Numbers}
\begin{center}
\begin{tabular}{|c||c|c|c|c|c|c|c|c|c|c|c|c|}
\hline
& \multicolumn{12}{c|}{${\rm U}(4)_C\times {\rm U}(2)_L\times {\rm
U}(2)_R \times {\rm U}(2) \times {\rm USp}(2)^2$}\\
\hline \hline  & $N$ & $(n^1,l^1)\times (n^2,l^2)\times

(n^3,l^3)$ & $n_{S}$& $n_{A}$ & $b$ & $b'$ & $c$ & $c'$& $d$ & $d'$ & 3 & 4 \\

\hline

     $a$&  8& $(0,-1)\times (1,1)\times (1,1)$ & \ 0 & \ 0  & 3 & 0 &
$-3$ & 0  & 0(2) & 0(1) & 0 & \ 0 \\

    $b$&  4& $(3,1)\times (1,~0)\times (1,-1)$ & \ 2 & $-2$  & - & - &
\ 0(6) & 0(1) & 1 & 0(1) & \ 0 & $-3$ \\

    $c$&  4& $(3,-1)\times (0,1)\times (1,-1)$ & $-2$ & \ 2  & - & - &
- & - & -1  & 0(1) &3 & \ 0 \\

    d&   4& $(1,0)\times (1,-1)\times (1,1)$ & 0 & 0 & - & - & - & - & - & - & -1 & 1
\\
\hline
    3&   2& $(0,-1)\times (~1,~0)\times (~0,~2)$ & \multicolumn{10}{c|}
{$\chi_1=3$}\\

    4&   2& $(0,-1)\times (~0,~1)\times (~2,0)$& \multicolumn{10}{c|}
{$\chi_2=1,~\chi_3=2$}\\

\hline

\end{tabular}

\end{center}

\end{table}

The set of D6 branes wrapping the cycles on a $T^6/(\Z_2 \times \Z_2)$
orientifold shown in Table~\ref{MI-Numbers} results in a
three-generation Pati-Salam model with additional hidden sectors.  The
full gauge symmetry of the model is given by $[{\rm U}(4)_C \times {\rm U}(2)_L \times {\rm
U}(2)_R]_{\rm observable} \times [ {\rm U}(2) \times {\rm USp}(2)^2]_{\rm hidden}$,
with the matter
content shown in Table~\ref{Spectrum}.  As discussed in detail
in~\cite{Chen:2007px,Chen:2007zu}, with this configuration of D6 branes all R-R
tadpoles are canceled, K-theory constraints are satisfied, and
$\mathcal{N}=1$ supersymmetry is preserved.  Furthermore, the
tree-level MSSM gauge couplings are unified at the string scale.
Finally, the Yukawa matrices for quarks and leptons are rank 3 and it
is possible to obtain correct mass hierarchies and mixings.  Note that
the observable sector of the model shown in Tables~\ref{MI-Numbers}
and~\ref{Spectrum} is identical to that of references~\cite{Chen:2007px,
Chen:2007zu} so that all of the above phenomenological features are also present.  However,
the hidden sector of the model is different, which as we shall see
gives rise to an extra anomaly-free U(1) gauge symmetry.

\begin{table}
[htb] \footnotesize
\renewcommand{\arraystretch}{1.0}
\caption{The chiral and vectorlike superfields, their multiplicities
and quantum numbers under the gauge symmetry $[{\rm U}(4)_C \times {\rm U}(2)_L \times {\rm
U}(2)_R]_{\rm observable} \times [ {\rm U}(2) \times {\rm USp}(2)^2]_{\rm hidden}$, where
$Q_X = Q_4 + 2(Q_{2L}+Q_{2R}+3Q_d)$.}
\label{Spectrum}
\begin{center}
\begin{tabular}{|c||c|c||c|c|c|c|c|c|c|c||c|}\hline
& Mult. & Quantum Number & $Q_4$ & $Q_{2L}$ & $Q_{2R}$ & $Q_X$ & Field
\\
\hline\hline
$ab$ & 3 & $(4,\overline{2},1,1,1,1)$ & \ 1 & $-1$ & \ 0 & $-1$ &
$F_L(Q_L, L_L)$\\
$ac$ & 3 & $(\overline{4},1,2,1,1,1)$ & $-1$ & 0 & \ 1  & \ 1 &
$F_R(Q_R, L_R)$\\
\hline
$bd$ & 1 & $(1,\overline{2},1,2,1,1)$ & 0 & $-1$ & 0  & $4$  & $X_{bd}$\\
$cd$ & 1 & $(1,1,2,\overline{2},1,1)$ & 0 & $0$ & 1  & $-4$  & $X_{cd}$\\
\hline
$b4$ & 3 & $(1,\overline{2},1,1,1,2)$ & 0 & $-1$ & 0  & $-2$  &
$X_{b3}^i$ \\
$c3$ & 3 & $(1,1,2,1,\overline{2},1)$ & 0 & 0 & \ 1  & \ 2 &
$X_{c3}^i$
\\
$d3$ & 1 & $(1,1,1,\overline{2},2,1)$ & 0 & $0$ & 1  & $-6$  & $X_{cd}$\\
$d4$ & 1 & $(1,1,1,2,1,\overline{2})$ & 0 & $0$ & 1  & $6$  & $X_{cd}$\\
$b_{S}$ & 2 & $(1,3,1,1,1,1)$ & 0 & \ 2 & 0   & \ 4 &  $T_L^i$ \\
$b_{A}$ & 2 & $(1,\overline{1},1,1,1,1)$ & 0 & $-2$ & 0  & $-4$ &
$S_L^i$
\\
$c_{S}$ & 2 & $(1,1,\overline{3},1,1,1)$ & 0 & 0 & $-2$  & $-4$ &
$T_R^i$
\\
$c_{A}$ & 2 & $(1,1,1,1,1,1)$ & 0 & 0 & \ 2  & 4 & $S_R^i$ \\
\hline\hline
$ab'$ & 3 & $(4,2,1,1,1,1)$ & \ 1 & \ 1 & 0  & \ 3 & $\Omega^i_L$ \\
& 3 & $(\overline{4},\overline{2},1,1,1,1)$ & $-1$ & $-1$ & 0 & $-3$ &
$\overline{\Omega}^i_L$ \\
\hline
$ac'$ & 3 & $(4,1,2,1,1,1)$ & \ 1 & 0 & \ 1  &  \ 3 & $\Phi_i$ \\
& 3 & $(\overline{4}, 1, \overline{2},1,1,1)$ & $-1$ & 0 & $-1$ & $-3$
& $\overline{\Phi}_i$\\
\hline
$bc$ & 6 & $(1,2,\overline{2},1,1,1)$ & 0 & 1 & $-1$  & \ 0 &
$H_u^i$, $H_d^i$\\
     & 6  & $(1,\overline{2},2,1,1,1)$ & 0 & $-1$ & 1  & \ 0 &
\\
\hline
\end{tabular}
\end{center}
\end{table}

Since ${\rm U}(N) = {\rm SU}(N)\times {\rm U}(1)$, associated with
each the stacks $a$, $b$, $c$, and $d$ are ${\rm U(1)}$ gauge groups,
denoted as ${\rm U}(1)_a$, ${\rm U}(1)_b$, ${\rm U(1)}_c$, and ${\rm U(1)}_d$.  
In general, these U(1)s are anomalous.  The anomalies associated with
these U(1)s are canceled by a generalized Green-Schwarz (G-S)
mechanism that involves untwisted R-R forms.  The couplings of the
four untwisted R-R forms $B^i_2$ to the U(1) field strength $F_a$ of
each stack $a$ are given by~\cite{CSU,Chen:2005aba}
\begin{eqnarray}
\label{GScouplings}
 N_a l^1_a n^2_a n^3_a \int_{M4}B^1_2\wedge \textrm{tr}F_a,  \;\;
 N_a n^1_a l^2_a n^3_a \int_{M4}B^2_2\wedge \textrm{tr}F_a \, ,
  \nonumber \\
 N_a n^1_a n^2_a l^3_a \int_{M4}B^3_2\wedge \textrm{tr}F_a,  \;\;
-N_a l^1_a l^2_a l^3_a \int_{M4}B^4_2\wedge \textrm{tr}F_a \, .
\end{eqnarray}
As a result, the gauge bosons of these Abelian groups generically
become massive.  However, these U(1)s remain as global symmetries to all
orders in perturbation theory. Indeed, baryon and lepton number conservation are typically
identified as arising from these global symmetries. 
These global U(1) symmetries
may also result in the forbidding of certain superpotential operators, such as Yukawa couplings
and those which mediate baryon and lepton number violation.
However, these {\it global} symmetries may be broken by nonperturbative effects, 
such as from D-brane instantons.  

The couplings of Eq.~(\ref{GScouplings}) determine the exact
linear combinations
of U(1) gauge bosons that acquire string-scale masses via the G-S
mechanism.  If U(1)$_X$ is a linear combination of the U(1)s from each
stack,
\begin{equation}
{\rm U}(1)_X \equiv \sum_a C_a {\rm U}(1)_a \, ,
\end{equation}
then the corresponding field strength must be orthogonal to those that
acquire G-S mass.  Thus, if a linear combination U(1)$_X$ 
satisfies~\cite{Chen:2005aba,Chen:2005mm,Chen:2005cf}
\begin{eqnarray}
&\sum_a C_a N_a l^1_a n^2_a n^3_a =0, \;\;
\sum_a C_a N_a n^1_a l^2_a n^3_a =0 \, ,
  \nonumber \\
&\sum_a C_a N_a n^1_a n^2_a l^3_a =0, \;\;
\sum_a C_a N_a l^1_a l^2_a l^3_a =0 \, ,
\label{GSeq}
\end{eqnarray}
the gauge boson of U(1)$_X$ acquires no G-S mass and is anomaly-free,
provided that the RR-tadpole conditions are satisfied.

For the present model, precisely one linear combination satisfies the
above conditions, and therefore has a massless gauge boson and is anomaly-
free:
\begin{equation}
{\rm U}(1)_X = {\rm U}(1)_a + 2\left[{\rm U}(1)_b + {\rm U}(1)_c + 3{\rm U}(1)_d
\right].
\end{equation}
Thus, the effective gauge symmetry of the model at the string scale is given by
\begin{equation}
{\rm SU}(4)_C \times {\rm SU}(2)_L \times {\rm SU}(2)_R \times {\rm
U}(1)_X \times \left[{\rm SU}(2) \times {\rm USp}(2)^2\right].
\end{equation}
As can be seen from Table~\ref{Spectrum}, the superfields $F_L^i(Q_L,
L_L)$ carry charge $Q_X = -1$, the superfields $F_R^i(Q_R, L_R)$ carry
charge $Q_X = +1$, while the Higgs superfields are uncharged under
U(1)$_X$.  Thus, the trilinear Yukawa couplings
for quarks and leptons are allowed by both the global
U(1) symmetries as well as the gauged ${\rm U}(1)_X$ symmetry.  As was
shown in~\cite{Chen:2007px,Chen:2007zu}, the resulting Yukawa matrices are rank
3, which allows for fermion mass textures that can easily accommodate
the observed mass hierarchies and mixings for quarks and leptons.  

The Pati-Salam gauge symmetry is broken to the SM in two steps.  
First, the $a$ and $c$ stacks of D6-branes are split such that
$a \rightarrow a1 + a2$ and $c \rightarrow c1 + c2$, where
$N_{a1}=6$, $N_{a2}=2$, $N_{c1}=2$, and $N_{c2}=2$.  The process
of breaking the gauge symmetry via brane splitting corresponds
to assigning VEVs along flat directions to adjoint scalars associated
with each stack
that arise from the open-string moduli~\cite{Cvetic:2004ui}.  
After splitting the D6-branes,
the gauge symmetry of the observable sector is 
\begin{equation}
SU(3)_C \times SU(2)_L \times U(1)_{I3R} \times U(1)_{B-L} \times U(1)_{3B+L},
\end{equation}
where 
\begin{eqnarray}
U(1)_{I3R} = \frac{1}{2}(U(1)_{c1} - U(1)_{c2}), \ \ \  
U(1)_{B-L} = \frac{1}{3}(U(1)_{a1} - 3U(1)_{a2}), 
\end{eqnarray}
and
\begin{equation}
U(1)_{3B+L}= -[U(1)_{a1}+U(1)_{a2} + 2(U(1)_{b}+U(1)_{c1}+U(1)_{c2}+3U(1)_{d})],
\end{equation}
and ${\rm U}(1)_{3B+L} = -{\rm U}(1)_X$.
Just as was the case in~\cite{Lebed:2011fw}, 
one may also form linear combinations of $U(1)_{B-L}$ and $U(1)_{3B+L}$ which
couple to baryon number and lepton number respectively:
\begin{eqnarray}
U(1)_B = \frac{1}{4}[U(1)_{B-L} + U(1)_{3B+L}], \ \ \ \ \ \ U(1)_L = \frac{1}{4}[-3U(1)_{B-L}+ U(1)_{3B+L}]. 
\label{GaugedBL}
\end{eqnarray}

As mentioned in the Introduction, the promotion of the SM to the MSSM
introduced operators which allow proton decay.  The first of
these is the rapid decay of the proton through the pair of $d =
4$ $F$-term operators 
($B$- and $L$-violating, respectively)~\cite{Ibanez:1991pr}:
\begin{equation}
\label{dim4operators}
U^c D^c D^c \, ,  \ Q D^c L \, .
\end{equation}
This problem is usually solved in the MSSM by introducing $R$ parity,
under which the known fermions are even while their SUSY partners are
odd (or the related ``matter parity'', under which $R = +1$ for $Q,
U^c \! , D^c \! , L, E^c \! , N^c$ and $R = -1$ for $H_{u,d}$).  As a
bonus, $R$ parity leads to a stable lightest SUSY particle (LSP),
which is a natural candidate for dark matter.  Although this idea is
attractive, it is well known that a gauged ${\rm U}(1)_{B-L}$ also
forbids the $d = 4$ operators, and furthermore, $R$ parity [more
specifically, matter parity $(-1)^{3(B-L)}$] can result from ${\rm
U}(1)_{B-L}$ broken spontaneously to its discrete $\Z_2$
subgroup~\cite{Font:1989ai, Krauss:1988zc, Martin:1992mq}.
For the present model, none of these operators are singlets
under either $U(1)_{B-L}$ or $U(1)_{3B+L}$ and so are 
forbidden.

Even though the problem of rapid proton decay via $d = 4$ operators
can be eliminated through this mechanism, one still faces the problem
of $d = 5$ operators that allow for proton decay with a lifetime too
short to evade current experimental constraints unless the
coefficients of these operators are chosen to be sufficiently small.
First among these are single operators that allow (at least in
principle) proton decay and preserve $B - L$:
\begin{equation}
\label{dim5operators}
[QQQL]_F \, , \ [U^c U^c D^c E^c]_F \, , \ [D^c D^c U^c N^c]_F \, .
\end{equation}
The second set consists of relevant $d = 5$ operators that violate
either $B$ or $L$ separately, which combine with the appropriate
member of Eq.~(\ref{dim4operators}) to form a composite operator that
conserves $B - L$ and allows proton decay:
\begin{eqnarray}
[QQQH_d]_F \, , \ [QU^cE^c H_d]_F \, , \ [QU^cL^{\dagger}]_D \, , \
[U^c (D^c)^{\dagger}E^c]_D \, , \ [QQ(D^c)^{\dagger}]_D \, ,
& & \nonumber \\
{} [QQ^{\dagger}N^c]_D \, , \ [U^c(U^c)^{\dagger}N^c]_D \, , \
[D^c(D^c)^{\dagger}N^c]_D \, , \ [QU^c N^c H_u]_F \, , \
[QD^c N^c H_d]_F \, . & &
\label{dim5operators2}
\end{eqnarray}
Indeed, these $d = 5$ operators are those which effectively lead to
the exclusion of GUTs based on minimal SU(5)~\cite{Murayama:2001ur},
although these operators can be suppressed in other unified models,
in particular flipped SU(5)~\cite{Barr:1981qv,Derendinger:1983aj,AEHN}.  For the
present model, it should be noted that these operators are invariant under 
${\rm U}(1)_{B-L}$; however
they are not invariant under ${\rm U}(1)_{3B+L}$. 
Thus, these operators are also forbidden in the model.  
Similar considerations apply to the dimension-6 proton decay operators. 
Of course, these results may be easily understood by considering that 
baryon and lepton number are effectively gauged in the model as given
by Eq. (\ref{GaugedBL}). 
It should also be emphasized that
since these operators are forbidden by gauged symmetries rather
than global symmetries, none of these operators may appear either 
perturbatively or nonperturbatively. 
Thus, the proton is essentially stable in this model with a lifetime
in excess of the current experimental lower bounds.  

Of course, the gauge symmetry must be further broken to the SM, with
the possibility of one or more additional U(1) gauge symmetries.  This 
may be accomplished in this model by assigning VEVs to the vectorlike
singlet fields with the quantum numbers $({\bf 1}, {\bf 1}, \frac 1 2,
-1, -3)$ and $({\bf 1}, {\bf 1}, -\frac 1 2, 1, 3)$ under the ${\rm
SU}(3)_C\times {\rm SU}(2)_L\times {\rm U}(1)_{I_{3R}}
\times {\rm U}(1)_{B-L} \times {\rm U}(1)_{3B+L}$ gauge symmetry from the
$a_2 c_2'$ intersections.  In this case, the gauge symmetry is further broken
to
\begin{equation}
\left[SU(3)_C \times SU(2)_L \times U(1)_Y \times U(1)_L\right]_{observable} \times \left[SU(2)\times USp(2)^2\right]_{hidden}, 
\end{equation}
where ${\rm U}(1)_L$ is given in Eq.~(\ref{GaugedBL}) and the electroweak hypercharge is given by
the combination
\begin{eqnarray}
&U(1)_Y = \frac{1}{6}\left[U(1)_{a1}-3U(1)_{a2}+3U(1)_{c1}-3U(1)_{c2}\right] \\ \nonumber
& = \frac{1}{2}U(1)_{B-L}+U(1)_{I3R}.
\end{eqnarray}
As we can see, if the gauge symmetry is broken to the SM in this way, $U(1)_L$ survives. 

On the other hand, other alternate scenarios for symmetry breaking are possible.  For
example, the U(1)$_{B-L}\times$U(1)$_{I_{3R}}\times$U(1)$_{3B+L}$
gauge symmetry may instead be broken by assigning VEVs to the
right-handed neutrino fields $N_R$.  In this case, the gauge symmetry
is broken to
\begin{equation}
\left[SU(3)_C \times SU(2)_L \times U(1)_Y \times U(1)_B\right]_{observable} \times \left[SU(2)\times USp(2)^2\right]_{hidden}. 
\label{MSSM_B}
\end{equation}
However, assigning VEVs to $N_R$ breaks SUSY, which is expected not to
occur until the TeV scale.  Thus, it is possible to obtain a nonanomlous gauged 
U(1) which counts either lepton number or baryon number, depending upon the 
way in which singlet VEVs are assigned.  

\section{Leptophobic and Leptophilic $Z'$ Bosons}

In the previous section we demonstrated that the gauge symmetry may be
broken to ${\rm SU}(3)_C \times {\rm SU}(2)_L \times {\rm U}(1)_Y
\times {\rm U}(1)_L$ at the GUT scale by assigning VEVs to the
vectorlike fields $\Phi$, $\overline{\Phi}$, or to ${\rm SU}(3)_C
\times {\rm SU}(2)_L \times {\rm U}(1)_Y \times {\rm U}(1)_B$ at the
TeV scale by assigning VEVs to the right-handed neutrinos $N_R$.
These two cases show that models of this type may be adapted to
provide either a U(1)$_L$ or U(1)$_B$ that survives unbroken to low
energies.  The possibility of constructing models where baryon and lepton
number are gauged at low energies has, of course, been considered
before~\cite{Foot:1989ts, FileviezPerez:2010gw, FileviezPerez:2011dg, FileviezPerez:2011pt}.
Usually in such models, extra matter must be arbitrarily added in order to
cancel anomalies.  For the present construction, the matter content and anomaly cancellation is fixed
by the configuration of D-branes and the global consistency conditions.  
Thus, it is possible to obtain nonanomlous U(1) gauge symmetries coupled to baryon and lepton number
in a very natural way (see~\cite{Cvetic:1995rj} for a discussion of the implications of extra Abelian gauge symmetries 
in string models). 

In addition to those fields discussed above, other singlet fields appear in the model whose VEVs may break
U(1)$_{3B+L}$ [or equivalently, U(1)$_B$ and U(1)$_L$] at intermediate
scales, namely, the singlets $S_L$ and $S_R$, as well as the SU(2)$_R$
triplet fields $T_R$.   
In particular, the $\mu$-term and a Majorana mass term may be generated by superpotential operators of the form 
\begin{eqnarray}
\label{mu-term}
W & \supset & \frac{y^{ijkl}_{\mu}}{M_{\rm St}} S_L^i S_R^j H_u^k H_d^l  \, 
+ \frac{y^{mnkl}_{Nij}}{M^3_{\rm St}} T_R^{m}
T_R^{n} \Phi^i \Phi^j  F_R^k  F_R^l ~,~\,
\end{eqnarray}
where $y^{ijkl}_{\mu}$ and $y^{mnkl}_{Nij}$ are Yukawa couplings.  
In this case, the
singlets $S_R$ and $T_R$ may obtain string or GUT-scale VEVs (or lower) 
while preserving the D-flatness of U(1)$_{2R}$,
and the singlets $S_L$ may obtain TeV-scale VEVs while preserving the
D-flatness of U(1)$_{2L}$, while the Higgses couple through their
electroweak-scale VEVs.
Simple order-of-magnitude estimates then show
that a TeV-scale $\mu$ term may be generated by these operators, with $y^{ijkl}_{\mu} =
O(1)$ and right-handed neutrino masses can be generated
in the range $10^{10-14}$~GeV for $y^{mnkl}_{Nij} \sim
10^{(-7)-(-3)}$, assuming GUT- or string-scale VEVs for the $\Phi$ and
$T_R$.  In this is the case, the the only surviving Abelian symmetry
in the model which survives is the SM hypercharge, U(1)$_Y$.  However, 
it is also possible to generate a $\mu$-term and a right-handed Majorana
mass via nonperturbative effects such as D-brane instantons~\cite{Blumenhagen:2009qh}.  
In this case, the singlet fields $S_R$ or $T_R$ need not receive VEVs at high
energy scales, and so either U(1)$_B$ or U(1)$_L$ may potentially survive unbroken.  

If either $U(1)_B$ or $U(1)_L$ survives unbroken down to the TeV-scale,
this may result in so-called leptophobic (coupled to quarks, but not leptons)
or leptophilic (coupled to leptons, but not quarks) $Z'$ 
bosons which may be observable at the LHC.  In particular, leptophobic 
Z' bosons have been obtained in unified models based on flipped SU(5) and E$_6$, though
typically with couplings which are family non-universal~\cite{Lopez:1996ta, Buckley:2011mm}.  
The possibility of observing
Z' bosons in general has been much studied in the literature and we direct the reader
to~\cite{Rizzo:2006nw} and~\cite{Langacker:2008yv} for reviews.

The main constraints on Z' bosons with electweak scale couplings come from
precision electroweak data, direct searches at the Tevatron, and searches 
for flavor changing neutral currents (FCNC). 
Perhaps the most stringent constraints on Z' couplings comes from LEP II.
For example, the process $e^+ e^- \rightarrow Z' \rightarrow e^+ e^-$ leads to a constraint of $g_{eeZ'} \lesssim 0.044 \times (m_{Z'}/200$~GeV for $Z'$ masses above roughly 200~GeV~\cite{lep:2003ih,Carena:2004xs,Bando:1987br}. At lower mass scales, the LEP II constraint, which is derived in an effective field theory formalism, is not directly applicable. Below about a scale of 200~GeV, off-shell $Z'$ production is not suppressed by the $Z'$ mass, but instead by the LEP center-of-mass energy. A modest constraint is therefore $g_{eeZ'} \lesssim 0.04$ for $m_{Z'} \lesssim 200$~GeV. Constraints which are somewhat strong may be placed on the production and decay into $e^+ e^-$ pairs of on-shell $Z'$ bosons if the $Z'$ mass is near one of the center-of-mass energies at which LEP~II operated~\cite{lep:2003ih}. Constraints from the $s$-channel production of $e^+ e^-$~\cite{Abulencia:2006xm} and/or $\mu^+ \mu^-$~\cite{Aaltonen:2008ah} at the Tevatron are also quite stringent ($\tau^+ \tau^-$ final states are considerably less constrained~\cite{Acosta:2005ij}). A $Z'$ with Standard Model-like couplings, for example, must be heavier than approximately 1~TeV to be consistent with the null results of these searches~\cite{Nakamura:2010zzi}.

 A so-called leptophobic $Z'$, such as would result from U(1)$_B$, is much more difficult to observe at both lepton and hadron colliders. In particular, at hadron colliders the QCD background at low dijet mass introduces large theoretical uncertainties, overwhelming any resonance signal arising from a $Z'$ with electroweak-strength or smaller couplings, thus the naive expectation that a search for a peak in the dijet invariant mass distributions would suffice is not correct. For a leptophobic $Z'$ in the mass range $\sim 300-900$~GeV, dijet searches at the Tevatron ($p \bar{p} \rightarrow Z' \rightarrow q \bar{q}$) constrain its couplings to quarks to be comparable to or less than those of the Standard Model $Z$ \cite{Aaltonen:2008dn}. For a leptophobic $Z'$ below 300 GeV, the uncertainties in the QCD background overwhelm the signal at the Tevatron, and so the strongest constraints come from the lower energy UA2 experiment~\cite{Alitti:1993pn}. From the lack of an observed dijet resonance,  UA2 can place constraints on the order of $g_{qqZ'} \lesssim 0.2$--$0.5$ for $Z'$ masses in the range of 130 to 300 GeV.
From these constraints, we can see that a leptophilic Z' resulting from U(1)$_L$ would require a mass greater than $1$~TeV, while a 
leptophobic Z' resulting from U(1)$_B$ may be light so long as its couplings to quarks are comparable to the Z boson of the SM. 

Both leptophilic and leptophobic Z' bosons have been put forward as explanations of various experimental anomalies in recent years. 
The possibility of leptophilic dark matter, such as might arise in the present context if 
the gaugino associated with U(1)$_L$ is stable,
has been suggested as an explanation~\cite{Fox:2008kb} of the observed
PAMELA~\cite{Adriani:2008zr}/ATIC~\cite{Chang:2008zzr} cosmic ray
positron excess, while a relatively light leptophobic 
$Z^\prime$ has been suggested as an explainion~\cite{Buckley:2011vc} of the
Tevatron anomalies in the measured $t \bar t$ forward-backward
asymmetry~\cite{Aaltonen:2011kc} and the associated production of $W$s
with jets~\cite{Aaltonen:2011mk}, although a more recent analysis by the CMS
collaboration has ruled out
a Z' as an explanation of the forward-backward
asymmetry~\cite{Chatrchyan:2011dk}. Furthermore, 
the D0 collaboration has not observed the same W +dijet excess as 
CDF~\cite{Abazov:2011af}.  The goal of the present work is not to provide a solution for these issues, 
but rather to demonstrate that such Z' bosons may exist in the model
and suggest possible applications (see~\cite{Anchordoqui:2011ag} and~\cite{Anchordoqui:2011eg} for
a similar recent discussion in the context of Type II string compactifications
with a low string scale).

As the possibility of low-scale Z' bosons has been extensively studied in the literarure, including leptophilic and leptophobic varieties, it is not necessary to repeat these analyses in the present context.  We have shown that the model may allow for such Z' bosons to be present at low-energies, and the results of previous studies on Z' bosons are applicable to these results. Most importantly, the Z' couplings in this model are family universal, thus they do not give rise to flavor-changing neutral currents (FCNC). Perhaps
the most exciting possibility for new physics involving Z' bosons at the moment is that a leptophobiz Z' can explain the W +dijet excess
reported by CDF~\cite{Buckley:2011vc, Hewett:2011nb}.   Finally, let us recall that the gauge symmetry of the may be broken so that U(1)$_B$ survives below the TeV scale by assigning VEVs to the right-handed neutrino fields,
$N_R$.  As the VEVs of these fields break supersymmetry, the scale at which this is expected is the TeV scale.  Thus, a leptophobic Z' in the model is expected
to have mass of the TeV scale or lower, while a leptophilic Z' may have a mass intermediate between the TeV scale and the unification scale.

\section{R-parity and Neutralino Dark Matter}

\begin{figure}[fth]
	\centering
		\includegraphics[width=0.75\textwidth]{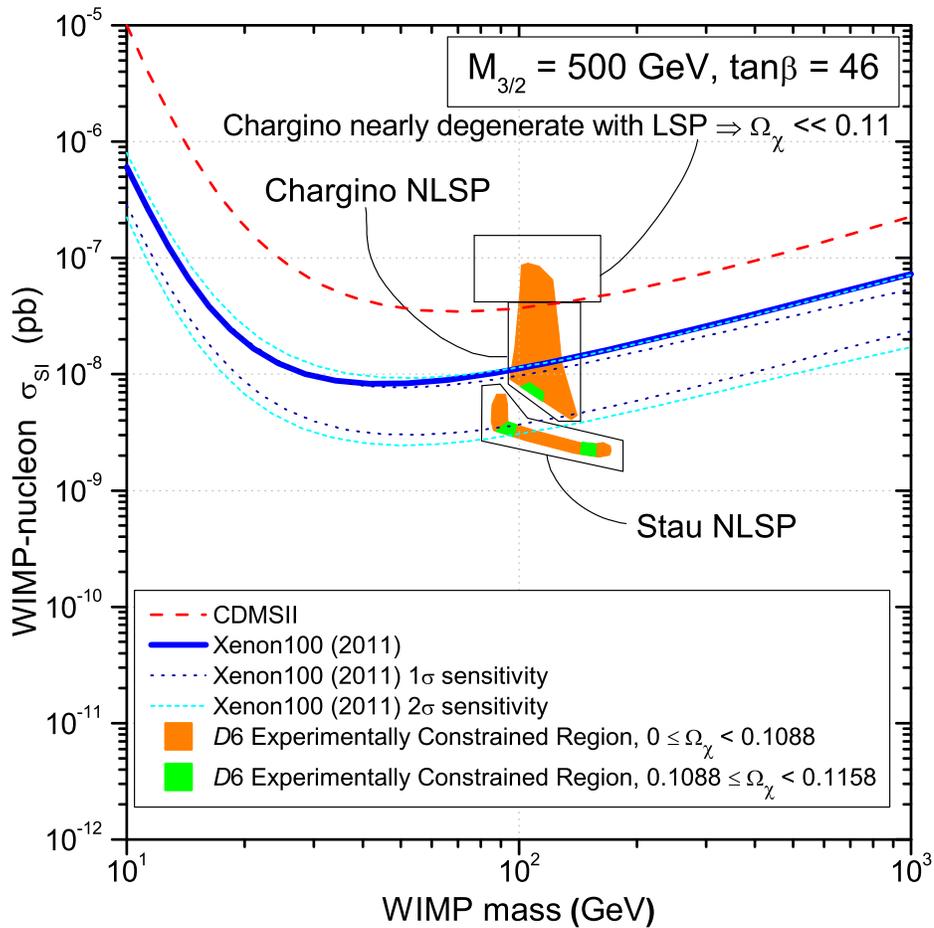}
		\caption{Direct dark matter detection diagram associating the WIMP mass with the spin-independent annihilation cross-section $\sigma_{\rm SI}$. Delineated are the current upper bounds from the CDMS~\cite{Ahmed:2009zw} and XENON100~\cite{Aprile:2011hx} experiments. Shown is the experimentally viable parameter space for a gravitino mass $M_{3/2}$ = 500 GeV and tan$\beta$ = 46. The boxes segregate the model space into the noted coannihilation regions.}
	\label{fig:sigma_plot1}
\end{figure}

\begin{figure}[fth]
	\centering
		\includegraphics[width=0.75\textwidth]{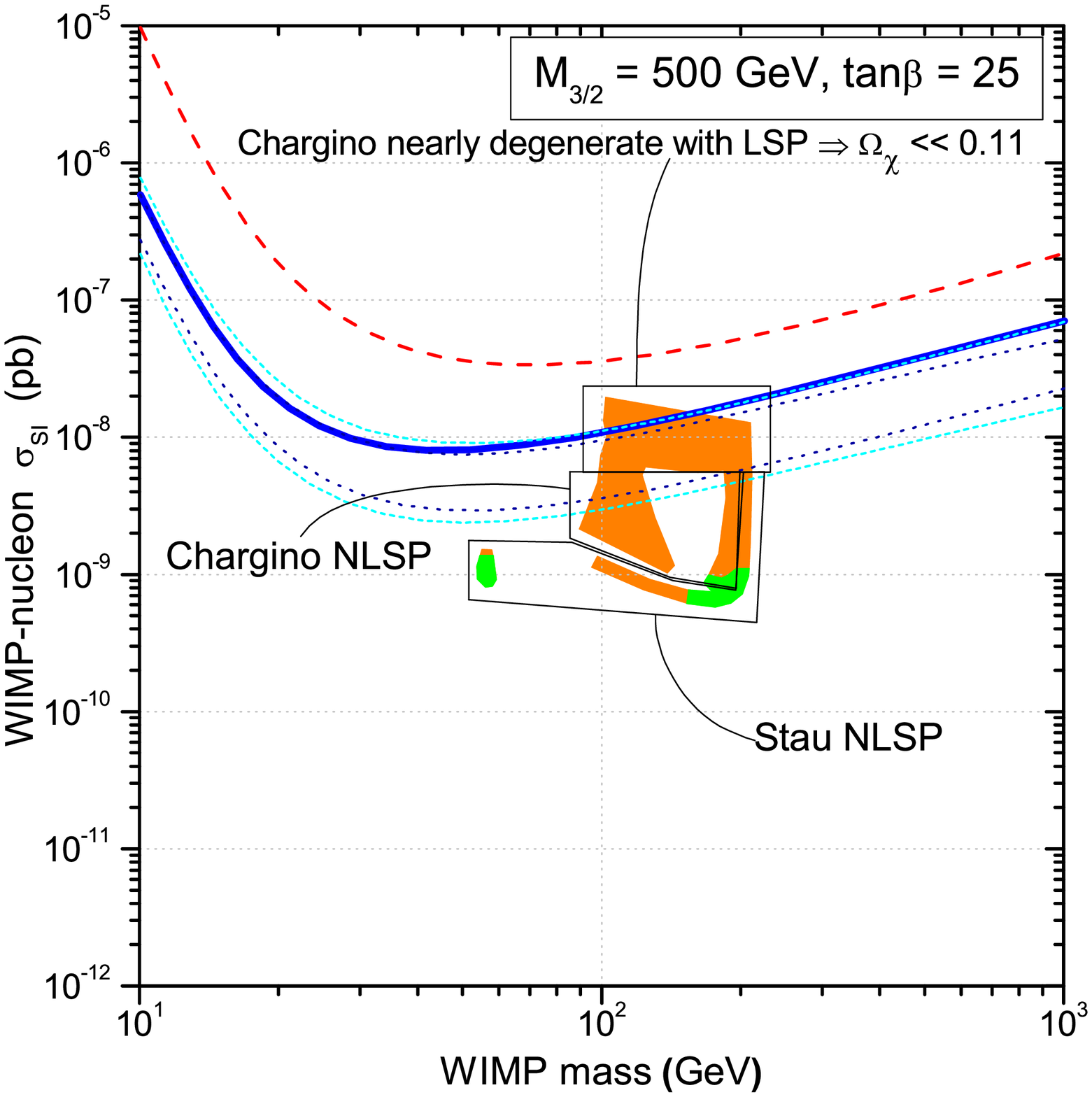}
		\caption{Direct dark matter detection diagram associating the WIMP mass with the spin-independent annihilation cross-section $\sigma_{\rm SI}$. Delineated are the current upper bounds from the CDMS~\cite{Ahmed:2009zw} and XENON100~\cite{Aprile:2011hx} experiments. Shown is the experimentally viable parameter space for a gravitino mass $M_{3/2}$ = 500 GeV and tan$\beta$ = 25. The boxes segregate the model space into the noted coannihilation regions. See Fig.~\ref{fig:sigma_plot1} for the legend describing the appropriate contours and regions.}
	\label{fig:sigma_plot2}
\end{figure}

\begin{figure}[fth]
	\centering
		\includegraphics[width=0.75\textwidth]{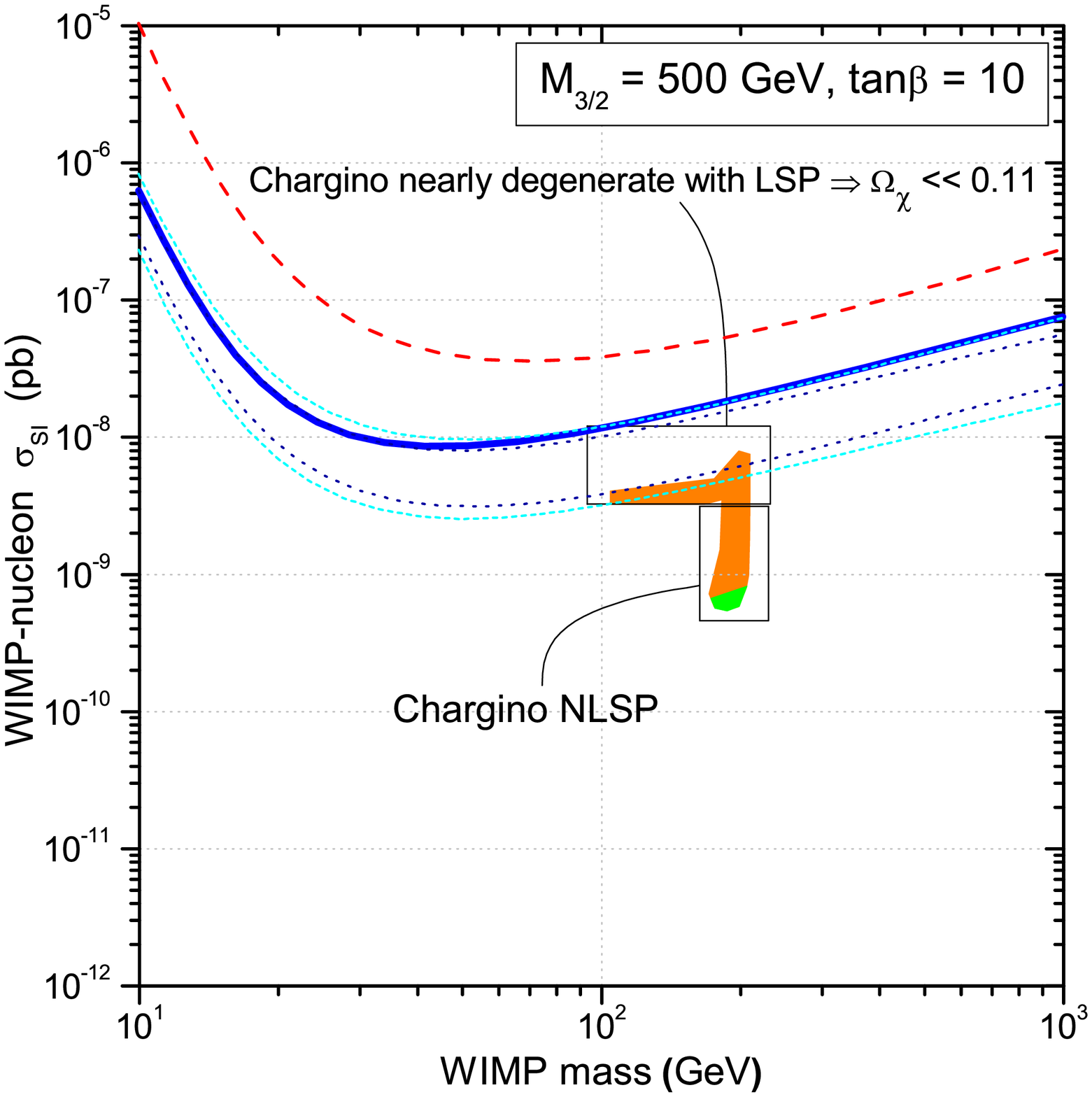}
		\caption{Direct dark matter detection diagram associating the WIMP mass with the spin-independent annihilation cross-section $\sigma_{\rm SI}$. Delineated are the current upper bounds from the CDMS~\cite{Ahmed:2009zw} and XENON100~\cite{Aprile:2011hx} experiments. Shown is the experimentally viable parameter space for a gravitino mass $M_{3/2}$ = 500 GeV and tan$\beta$ = 10. The boxes segregate the model space into the noted coannihilation regions. See Fig.~\ref{fig:sigma_plot1} for the legend describing the appropriate contours and regions.}
	\label{fig:sigma_plot3}
\end{figure}

\begin{figure}[fth]
	\centering
		\includegraphics[width=0.75\textwidth]{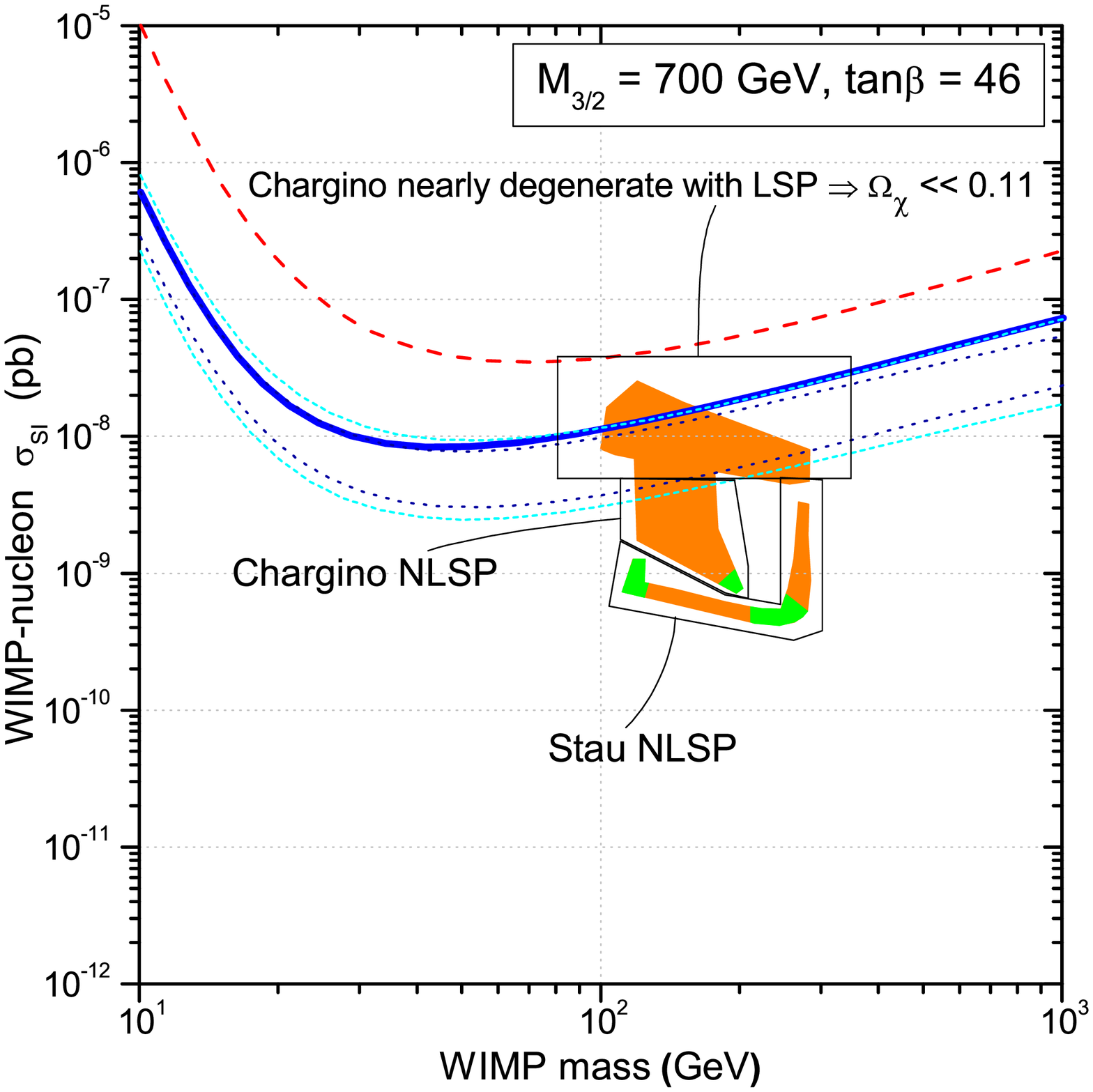}
		\caption{Direct dark matter detection diagram associating the WIMP mass with the spin-independent annihilation cross-section $\sigma_{\rm SI}$. Delineated are the current upper bounds from the CDMS~\cite{Ahmed:2009zw} and XENON100~\cite{Aprile:2011hx} experiments. Shown is the experimentally viable parameter space for a gravitino mass $M_{3/2}$ = 700 GeV and tan$\beta$ = 46. The boxes segregate the model space into the noted coannihilation regions. See Fig.~\ref{fig:sigma_plot1} for the legend describing the appropriate contours and regions.}
	\label{fig:sigma_plot4}
\end{figure}

\begin{figure}[fth]
	\centering
		\includegraphics[width=0.75\textwidth]{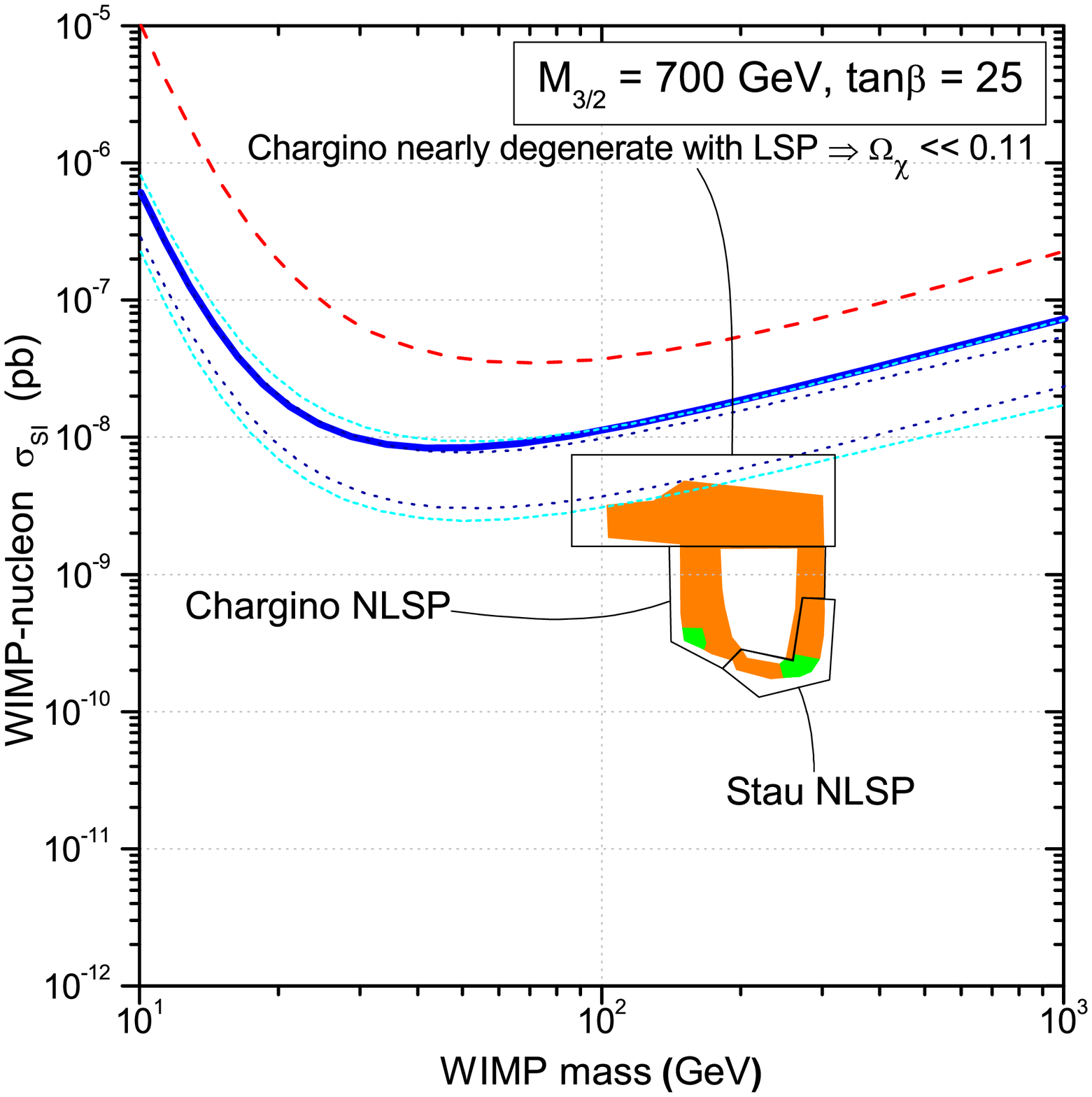}
		\caption{Direct dark matter detection diagram associating the WIMP mass with the spin-independent annihilation cross-section $\sigma_{\rm SI}$. Delineated are the current upper bounds from the CDMS~\cite{Ahmed:2009zw} and XENON100~\cite{Aprile:2011hx} experiments. Shown is the experimentally viable parameter space for a gravitino mass $M_{3/2}$ = 700 GeV and tan$\beta$ = 25. The boxes segregate the model space into the noted coannihilation regions. See Fig.~\ref{fig:sigma_plot1} for the legend describing the appropriate contours and regions.}
	\label{fig:sigma_plot5}
\end{figure}

As discussed in the previous section, all dimension-4, 5, and 6 operators which arise
in the MSSM that may mediate proton decay are forbidden in this model by the extra
U(1) gauge symmetry.  In the conventional MSSM, the dimension-4 operators which lead 
to proton decay at a disastrously high rate are typically removed by invoking R-parity.
As a bonus, this results in a stable LSP, which can provide an excellent dark matter candidate
in the case of a neutralino or gravitino LSP.  However, as we have seen, it is not
necessary to invoke R-parity in this model in order eliminate rapid proton decay.  
Thus, it is possible that the LSP may not be stable in this model, and therefore would
not provide a dark matter candidate.  However, the model does possess a gauged
${\rm U}(1)_{B-L}$ at the string scale after the Pati-Salam gauge symmetry is broken.  
This can then provide a natural origin for a gauged R-parity, even though it is not required for 
proton stability. In particular, the model will possess an exact gauged R-parity provided that 
${\rm U}(1)_{B-L}$ is broken by scalar VEVs that carry even integer values of $3(B-L)$~\cite{Martin:1992mq}.  
As this can clearly be accomplished in the model, in the following we will consider that 
an exact gauged R-parity does exists and a stable LSP provides the required dark matter candidate.  
We then will 
analyze the constraints on possible cross-sections for direct dark matter detection in 
light of the latest experimental data from the XENON100 and CDMS experiments.  However, it should be 
noted that the case without exact R-parity would also be very interesting to study.  In
particular, this could result in different decay cascades as well as the 
absence of large missing energy signals since the LSP would not be stable.  Needless to say,
such a scenario could make it somewhat more difficult to observe superpartners at the LHC.  For this 
reason, as well as others, it is therefore very important to also study dark matter direct-detection 
experiments in order to compare the predictions of supersymmetric models with the actual
properties of the dark matter.  

In contrast to phenomenological frameworks such as mSUGRA, the supersymmetry
breaking soft terms in intersecting $D6$-branes are
in general non-universal~\cite{Kors:2003wf}. Thus, it is possible to obtain
a parameter space which is more general than in mSUGRA. A detailed discussion 
of the supersymmetry parameter space of the $D6$ model may be found 
in~\cite{Chen:2007px, Chen:2007zu,Maxin:2009ez}. A comparison of the superpartner parameter
space for the present case with nonuniversal soft-terms to one-parameter models motivated by
no-scale supergravity may be found in~\cite{Maxin:2008kp,Maxin:2009pr}. The low-energy effective
action for intersecting D-brane models has been given in~\cite{Brignole:1997dp,Kors:2003wf,Lust:2004cx}
while explicit formulas for the soft-supersymmetry 
breaking terms used to generate the phenomenology in this work are contained in 
Ref.~\cite{Kane:2004hm,Maxin:2009ez}. We assume that the gauge symmetry of the observable sector
consist of only the MSSM below the usual GUT scale, $M_{GUT}=2.2\cdot10^{16}$~GeV,
hence the observable supersymmetric phenomenology should 
remain consistent with formulae of Ref.~\cite{Maxin:2009ez}. To examine the 
dark matter content of the $D6$ model space, we investigate regions of the 
intersecting $D$6-brane model parameter space that satisfy all of the most 
current experimental constraints.  The soft terms are input 
into {\tt MicrOMEGAs 2.0.7}~\cite{Belanger:2006is} using {\tt SuSpect 2.34}~\cite{Djouadi:2002ze} 
as a front end to run the soft terms down to the 
electroweak scale via the Renormalization Group Equations (RGEs) and then to 
calculate the corresponding relic neutralino density, while $\mu$ is determined 
by the requirement of radiative electroweak symmetry breaking (REWSB). However, 
we do take $\mu > 0$ as suggested by the results of $g_{\mu}-2$ for the muon. We 
use the current world average central value top quark mass of $m_{t}$ = 173.1 GeV~\cite{:2009ec}. 
The direct detection cross-sections are calculated using {\tt MicrOMEGAs 2.1}~\cite{Belanger:2008sj}. 
We apply the following experimental constraints:

\begin{enumerate}

\item The 7-year WMAP measurements of the cold dark matter 
density~\cite{Komatsu:2010fb}, 0.1088 $\leq \Omega_{\chi} \leq$ 0.1158. 
We also investigate another case where a neutralino LSP makes up a 
subdominant component and employ this possibility by removing the lower bound.

\item The experimental limits on the Flavor Changing Neutral Current (FCNC) 
process, $b \rightarrow s\gamma$. The results from the Heavy Flavor Averaging 
Group (HFAG)~\cite{Barberio:2007cr}, in addition to the BABAR, Belle, and 
CLEO results, are: $Br(b \rightarrow s\gamma) = (355 \pm 24^{+9}_{-10} \pm 3) \times 10^{-6}$. 
There is also a more recent estimate~\cite{Misiak:2006zs} of $Br(b \rightarrow s\gamma) = (3.15 \pm 0.23) \times 10^{-4}$. For our analysis, we use the limits $2.86 \times 10^{-4} \leq Br(b \rightarrow s\gamma) \leq 4.18 \times 10^{-4}$, 
where experimental and theoretical errors are added in quadrature.

\item The anomalous magnetic moment of the muon, $g_{\mu} - 2$. For this 
analysis we use the 2$\sigma$ level boundaries, $11 \times 10^{-10} < a_{\mu} < 44 \times 10^{-10}$~\cite{Bennett:2004pv}.

\item The process $B_{s}^{0} \rightarrow \mu^+ \mu^-$ where the decay 
has a $\mbox{tan}^6\beta$ dependence. We take the upper bound to 
be $Br(B_{s}^{0} \rightarrow \mu^{+}\mu^{-}) < 5.8 \times 10^{-8}$~\cite{:2007kv}.

\item The LEP limit on the lightest CP-even Higgs boson mass, $m_{h} \geq 114$ GeV~\cite{Barate:2003sz}.

\end{enumerate}

We present the updated WIMP-nucleon spin-independent cross-section contours 
in Figs.~\ref{fig:sigma_plot1}-\ref{fig:sigma_plot5}. Following the methodology 
in Ref.~\cite{Maxin:2009ez}, we segregate the parameter space into distinctive 
scenarios of $m_{3/2}$ and tan$\beta$. The five scenarios of $m_{3/2}$ and 
tan$\beta$ first introduced in~\cite{Maxin:2009ez} and also analyzed here in 
this work were selected to be representative of a broad range of the experimentally 
allowed parameter space. To satisfy the LEP limit on the lightest CP-even Higgs 
boson mass, the gravitino mass needs to generally be a minimum of about 
$m_{3/2}$ = 500 GeV, while the observability of the $D6$ model at the 
near-term LHC is questionable for a gravitino mass greater than $m_{3/2}$ = 700 GeV 
due to very heavy superpartners, hence the range of $m_{3/2}$ used for these 
analyses. Likewise, the tan$\beta$ examined here are those from near the minimum 
tan$\beta$ possible to satisfy the experimental constraints, to a large tan$\beta$ 
value representative of the high tan$\beta$ region of the model space. The spin-independent 
contours in Figs.~\ref{fig:sigma_plot1}-\ref{fig:sigma_plot5} represent the most current 
upper bounds from the CDMS~\cite{Ahmed:2009zw} and XENON100~\cite{Aprile:2011hx} experiments, 
as a function of the LSP mass. We find that the only regions significantly affected by the 
XENON100 constraints are those where the lightest neutralino and chargino are 
nearly degenerate. These points of chargino-neutralino degeneracy possess a nearly 
zero relic density, thus the neutralino would comprise only a tiny fraction of the total 
cold dark matter. Nonetheless, those regions with both chargino-neutralino and 
stau-neutralino coannihilation do subsist for potential supersymmetry and LSP discovery. 
Each figure is demarcated to clearly identify the appropriate areas of coannihilation.

\section{Conclusion}

We have discussed the presence of an extra nonanomalous U(1) gauge symmetry in a 
realistic three-generation Pati-Salam model constructed with intersecting D6-branes in 
Type IIA string theory on a $T^6/(\Z_2 \times \Z_2)$ orientifold.  As discussed in previous
papers, the SM gauge couplings are unified at the string scale in this model and it is 
possible to obtain realistic Yukawa matrices for quarks and leptons.  Besides these favorable 
phenomenological features, we have shown that the additional 
U(1) prohibits all dimension-4, 5, and 6 operators that mediate 
proton decay in the MSSM. In particular, we have shown that this U(1) gives rise to a 
${\rm U}(1)_{3B+L}$ gauge symmetry once the Pati-Salam gauge symmetry is broken.  Furthermore, 
it is possible to find linear
combinations with ${\rm U}(1)_{B-L}$ which lead to the effective promotion of baryon and lepton
number to local gauge symmetries.  Thus, the proton is essentially stable in this model.  

As was mentioned, in the MSSM rapid proton decay is allowed by the dimension-4 operators unless
a discrete symmetry such as R-parity is imposed. 
However, even though these operators can be removed without invoking R-parity in the model considered here, R-parity 
can naturally surface from a spontaneously broken $U(1)_{B-L}$ provided certain conditions are satisfied
which can be accomplished in the model. Thus, presuming the existence
of R-parity giving rise to a stable LSP, we studied the direct detection cross-sections for neutralino dark matter 
through application of the most current constraints from the XENON100 and CDMS experiments. We 
found that other than those regions with a lightest neutralino and chargino degeneracy, 
the $D6$ model space remains relatively intact and unaffected by the XENON100 constraints. 
These include regions of the parameter space where the relic density is generated through 
neutralino coannihilation with the stau and lightest chargino.  However, this experiment will soon
have sufficient reach to thoroughly test the model predictions for stable neutralino dark matter.  

We should comment that it is really quite remarkable that a nonanomalous U(1) gauge symmetry 
arises in the model which automatically leads to a stable proton with a very long lifetime.  
Given the observed long lifetime of the proton, 
proton stability can be considered one of the essential properties of any model of particle physics,
string theory vacua in particular, given the existence of the String Landscape.  
It is known that models built on minimal ${\rm SU}(5)$ do not satisfy this
criteria as the dimension-5 operators which mediate proton decay are present in such constructions.  However,
in unified models based on ${\rm SU}(5)$ such as flipped SU(5), the proton may decay
at a rate which is observable, but which satisfies current experimental limits.  Although the model considered
in this paper has a Pati-Salam structure, we have pointed out that it is not possible to obtain this Pati-Salam
from a GUT such as ${\rm SO}(10)$ due to the charges under the extra U(1) carried by the matter supermultiplets.  Thus,
this scenario can be considered to give rise to a new \lq GUT-less\rq \ paradigm where the proton is stable
and the gauge couplings are unified at high energies, but where the gauge symmetry does not unify to a grand theory 
(other than of the Pati-Salam type).     

\section{Acknowledgments}

This was supported in part by the National Science Foundation
under Grant No.\ PHY-0757394 (VEM), and by the Mitchell-Heep Chair in High Energy Physics and DOE grant DE-FG03-95-Er-40917 (DVN and JAM).

\end{document}